\documentclass[11pt]{article}
\usepackage{moriond,epsfig}
\usepackage{xspace}
\usepackage{graphicx}
\usepackage{amsmath,amssymb,amsfonts}
\usepackage[english]{babel}

\def\Journal#1#2#3#4{{#1} {\bf #2}, #3 (#4)}
\def\Title#1{{\it #1}}
\def\Conf#1#2#3#4{{#1}, {\it #2}, {#3}, {#4}}

\def\ZPC{{\em Z. Phys.} C}

\def\JPG{{\em J. Phys.} G}
\def\EPJC{{\em Eur. Phys. J.} C}
\def\JH{{\em JHEP}}

\def\be{\begin{equation}}
\def\ee{\end{equation}}
\def\bea{\begin{eqnarray}}
\def\eea{\end{eqnarray}}

\newcommand{\coll}{Coll.\xspace}
\newcommand{\etal}{et al.\xspace}

\newcommand{\ddif}[3]{\frac{d^{2}#1}{d#2 d#3}}
\newcommand{\eVdist}{\kern-0.06667em}

\newcommand{\gev}{{\,\text{Ge}\eVdist\text{V\/}}}

\newcommand{\pbi}{\,\text{pb}^{-1}}

\begin{document}
\vspace*{4cm}
\title{PROTON STRUCTURE FUNCTIONS AT HIGH $Q^{2}$ AND HIGH $x$ AT HERA}

\author{ S.~U.~NOOR \\ (On behalf of the ZEUS and H1 collaborations) }

\address{York University, \\
Petrie Science and Engineering Building,  \\
4700 Keele St., Toronto, 
Ontario, M3J 1P3, Canada
}
\maketitle
\abstracts{
Proton structure measurements at high $Q^{2}$ and high $x$, performed by the H1
and ZEUS collaborations at the HERA collider, are reviewed. Neutral and
charged current deep inelastic scattering cross sections and structure
functions are presented. 
The review also discusses improvements to the parton density measurements
using jet cross section data and recent high $Q^{2}$ inclusive cross section
measurements. 
The projected parton density
uncertainties using  the entire HERA data set are also presented.} 
\section{Introduction}
Precise measurements of the proton parton density functions (PDFs) are crucial
for understanding 
the structure of the proton. This is of particular importance 
with the imminent start of the LHC proton-proton collider.
Unpolarised lepton beams were used before the luminosity
upgrade in 2000 (HERA-I), whereas the post-upgrade
collider (HERA-II) has delivered polarised leptons beams.
This paper reviews the latest measurements at high $Q^{2}$
and high $x$ performed by the ZEUS and H1 collaborations at HERA.

The PDFs are determined in global fits at next-to-leading order (NLO)
in QCD using data from deep inelastic scattering (DIS) experiments. The 
kinematic range covered by HERA has allowed the determination of PDFs across a
wide range of phase space spanned by the fractional proton momentum of the
struck quark, Bjorken-$x$, and the negative squared four-momentum transfer,
$Q^{2}$, of approximately $10^{-6} < x < 1$ and $0 <
Q^{2} < 10^{5} \gev^{2}$.   

\section{Cross Section Measurements and Structure Functions}
The Born-level reduced cross section for the $e^ \pm p$
neutral current (NC) interaction with polarised lepton beams can be written
as~\cite{dev}, 
\begin{equation}
  \tilde{\sigma}^{e^{\pm} p} 
  =
  \frac {xQ^{4}} {2 \pi \alpha^{2} }
  \frac {1} {Y_{+}}
  \ddif{\sigma(e^{\pm}p)}{x}{Q^{2}}
  =
  F^{\pm}_{2}(x,Q^{2}) \mp \frac {Y_{-}} {Y_{+}} xF^{\pm}_{3}(x,Q^{2})
  - \frac {y^{2}} {Y_{+}} F^{\pm}_{L}(x,Q^{2}), 
\label{eqn:red}
\end{equation}
where  $\alpha$ is the fine-structure constant, $Y_{\pm} \equiv 1 \pm (1 -
y)^{2}$ and $y$ is related to the centre-of-mass energy, $\sqrt{s}$, via
$Q^{2}=sxy$. The longitudinal structure function, $F_{L}$,  
is small in the kinematic region considered and can be ignored.    
The structure functions, $F_{2}$ and $xF_{3}$, contain the sum and difference
of the quark and anti-quark PDFs and can be separated into
contributions from pure $\gamma$ exchange, the interference of $\gamma$ and
$Z$ boson exchange and from pure $Z$ exchange. 
These terms depend on the lepton beam charge, the longitudinal
polarisation of the lepton beam, $P_{e}$, the mass of the $Z$ and $W$ bosons,
$M_{Z}$ and $M_{W}$, and the weak-mixing angle, $\theta$, to give the
following~\cite{klein}, 
\be
F_{2}^{\pm} = F^{\gamma}_{2} + k(-v_{e} \mp P_{e}a_{e})F^{\gamma Z}_{2} +
k^{2}(v^{2}_{e} + a^{2}_{e} \pm 2P_{e}v_{e}a_{e})F^{Z}_{2},
\ee
\be
xF^{\pm}_{3} = k(-a_{e} \mp P_{e}v_{e})xF^{\gamma Z}_{3} + 
k^{2}(2v_{e}a_{e} \pm P_{e}(v^{2}_{e} + a^{2}_{e}))xF^{Z}_{3},
\ee
where $k = \frac{1}{4\sin^{2}\theta \cos^{2}\theta}\frac{Q^{2}}{Q^{2} +
  M^{2}_{Z}}$ and the vector and axial-vector coupling of the electron to the
$Z$ boson are $v_{e} = -1/2 + 2\sin^{2}\theta$ and $a_{e} = -1/2$
respectively.
\begin{figure}[t]
  \begin{center}
    $\begin{array}{c@{\hspace{0.3in}}c}
      \epsfxsize=2.75in
      \epsffile{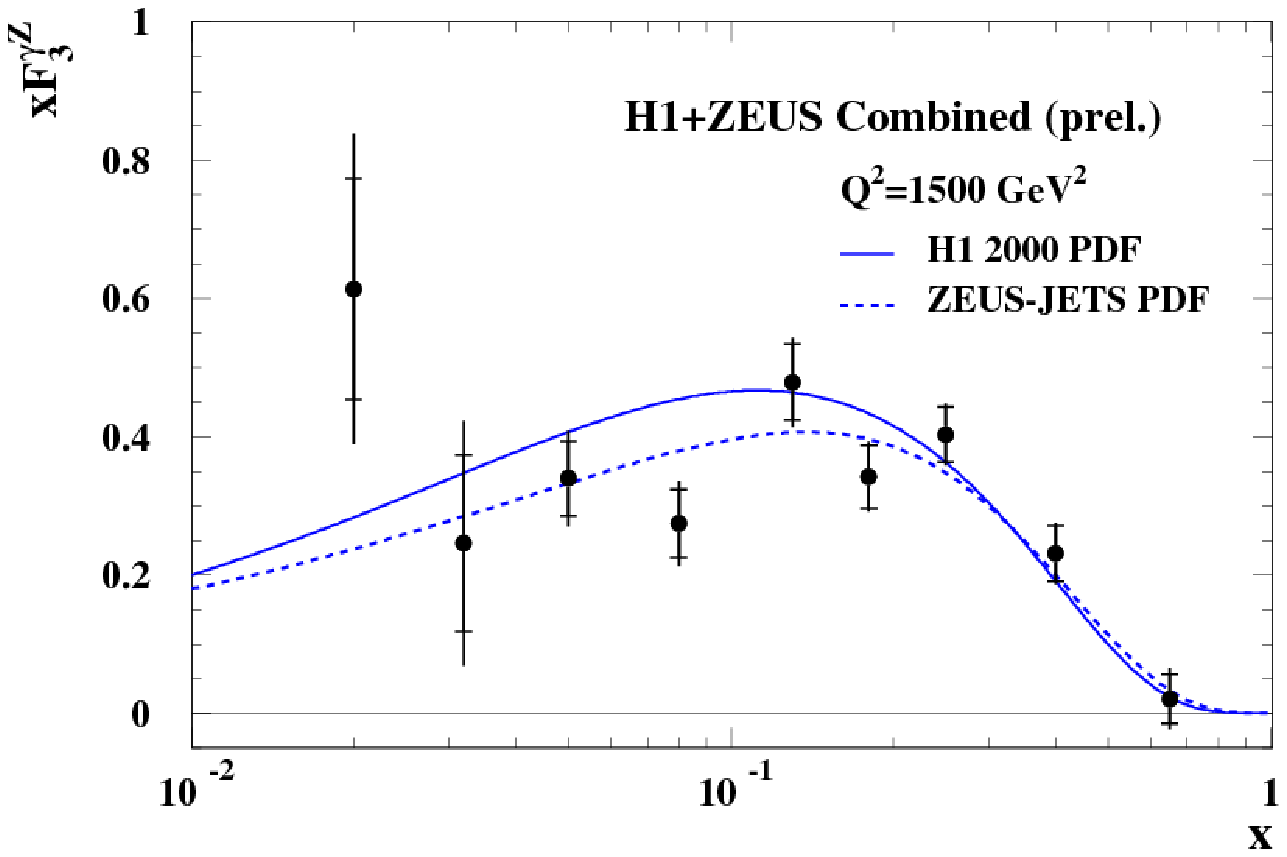} &
      \epsfxsize=2.75in
      \epsffile{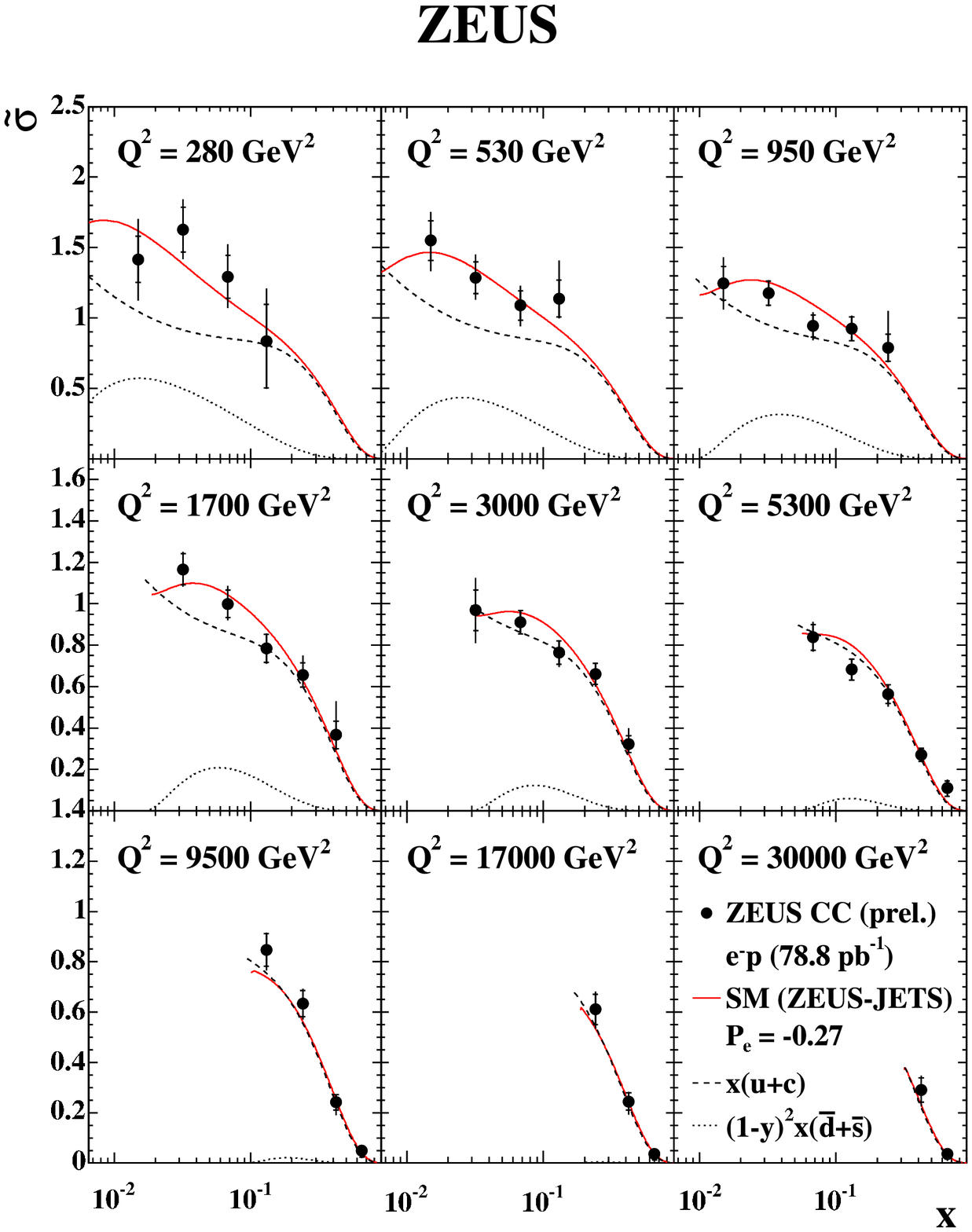} \\
      (a) & (b) \\ [0.4cm]
    \end{array}$
  \end{center}
  \vspace{-0.5cm}
  \caption{  Graph (a) shows the combined measurement of the structure
    function $xF_{3}^{\gamma  Z}$ versus $x$ at  $Q^{2}=1500 \gev^{2}$ from
    the H1 and ZEUS collaborations using $e^{\pm}p$ NC DIS data. 
    The curves represent the SM
    prediction from  the H1 2000 and ZEUS-JETS PDFs.
    The plots shown in (b) present the CC $e^{-}p$ reduced cross section as a
    function of
    $x$ in fixed bins of $Q^{2}$. The SM prediction using the ZEUS-JETS PDF is
    shown in red and dashed lines indicate the contributions from terms
    involving up type quarks and down type anti-quarks. 
 } 
  \label{fig-xf3}
\end{figure}
By taking the difference $\tilde{\sigma}^{e^{-}p}- \tilde{\sigma}^{e^{+}p}$ 
one can extract $xF_{3}$ using unpolarised HERA-I data and net unpolarised
data from HERA-II. 
As the polarisation dependence is removed, $xF_{3}$ can be written as,
\be
xF_{3} = -a_{e} k xF^{\gamma  Z}_{3} +
2v_{e}a_{e} k ^{2}xF^{Z}_{3}.
\ee
Since the coupling $v_{e}$ is small and $ k  < 1$, the interference term
dominates $xF_{3}$. In leading order (LO) perturbative QCD the interference
structure function can be explicitly written in terms of the valence quark
distributions, $u_{v}$ and $d_{v}$, 
\be
xF^{\gamma Z}_{3} = \frac{x}{3}(2u_{v} + d_{v} + \Delta),
\ee
where $\Delta = 2(u_{sea} -\bar{u} + c - \bar{c}) + (d_{sea} - \bar{d} + s -
\bar{s})$. 
Therefore $xF^{\gamma Z}_{3}$ 
is determined by the valence quark distribution if the $\Delta$ term is
ignored, and is only weakly dependent on $Q^{2}$. 
To minimise statistical errors, the $xF^{\gamma Z}_{3}$ measurements 
can be extrapolated in $Q^{2}$ and averaged in $x$. Results from the ZEUS and H1
collaborations are shown in Fig.~\ref{fig-xf3}(a).  

The Born level charged current (CC) $e^{\pm}p$ cross
section with polarised leptons can be expressed at LO in QCD as~\cite{dev},
\be
\ddif{\sigma_{CC}(e^{-}p)}{x}{Q^{2}} = 
(1-P_{e})
\frac{G_{F}^2}{2\pi}
\left( \frac{M^{2}_{W}}{M^{2}_{W} + Q^{2}} \right)^{2}
[u+c+(1-y)^2(\bar{d}+\bar{s})],
\ee
\be
\ddif{\sigma_{CC}(e^{+}p)}{x}{Q^{2}} = 
(1+P_{e})
\frac{G_{F}^2}{2\pi}
\left( \frac{M^{2}_{W}}{M^{2}_{W} + Q^{2}} \right)^{2}
[\bar{u}+\bar{c}+(1-y)^2(d+s)],
\ee
where $G_{F}$ is the Fermi coupling constant and $u, c, d, s$ are the
respective quark 
densities. The flavour selecting nature of the CC interaction is apparent as 
$u$ quark content is revealed through $e^{-}p$ DIS, whereas $d$ quark
constraints are possible through $e^{+}p$ scattering.
This can be illustrated in the $e^{-}p$ CC DIS reduced cross
section measurements~\cite{zcc} shown in Fig.~\ref{fig-xf3}(b), where the SM prediction
describes the data well and is dominated by the $u$ quark density. 
\begin{figure}[t]
  \begin{center}
    $\begin{array}{c@{\hspace{0.3in}}c}
      \epsfxsize=2.75in
      \epsffile{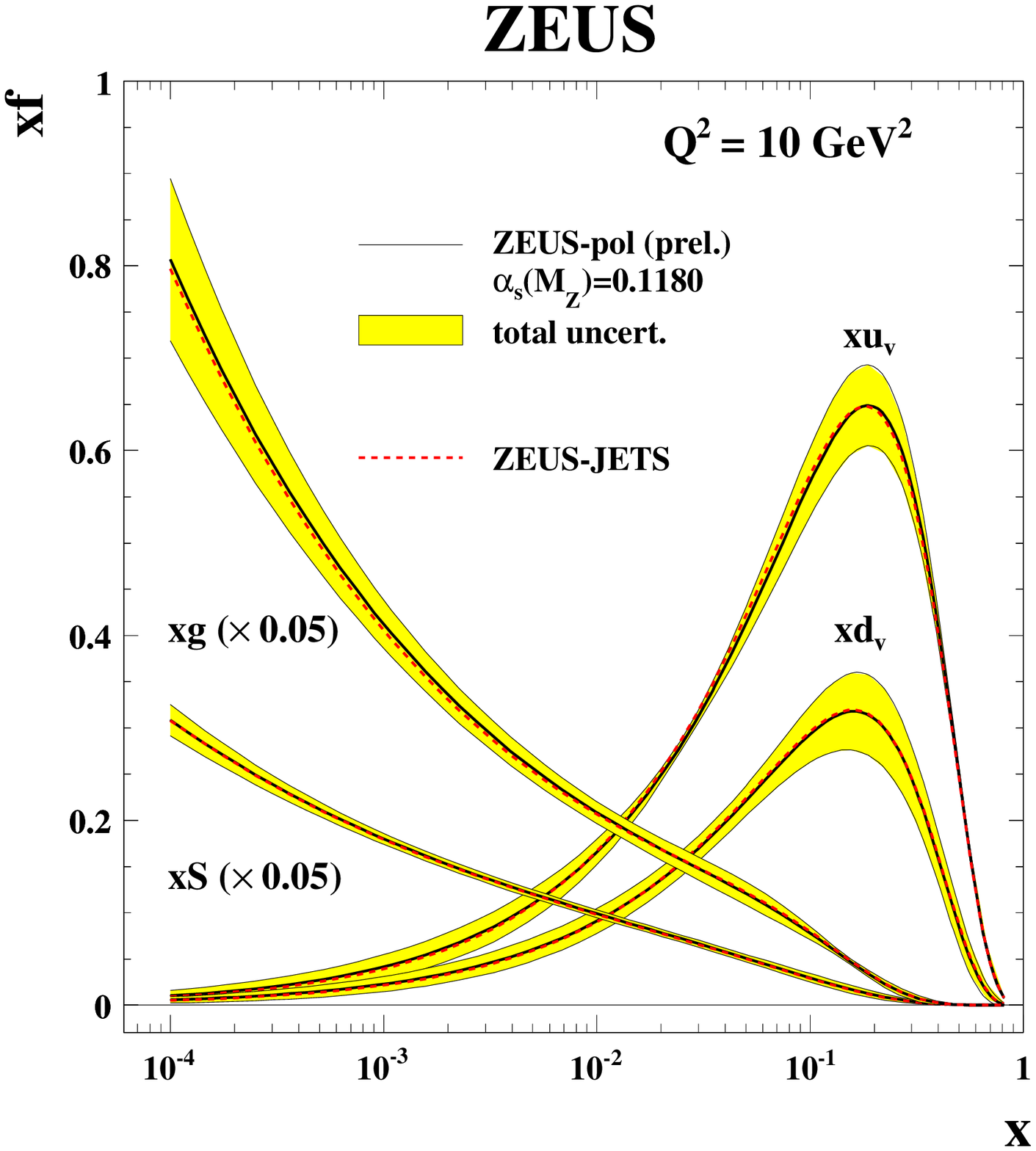} &
      \epsfxsize=2.75in
      \epsffile{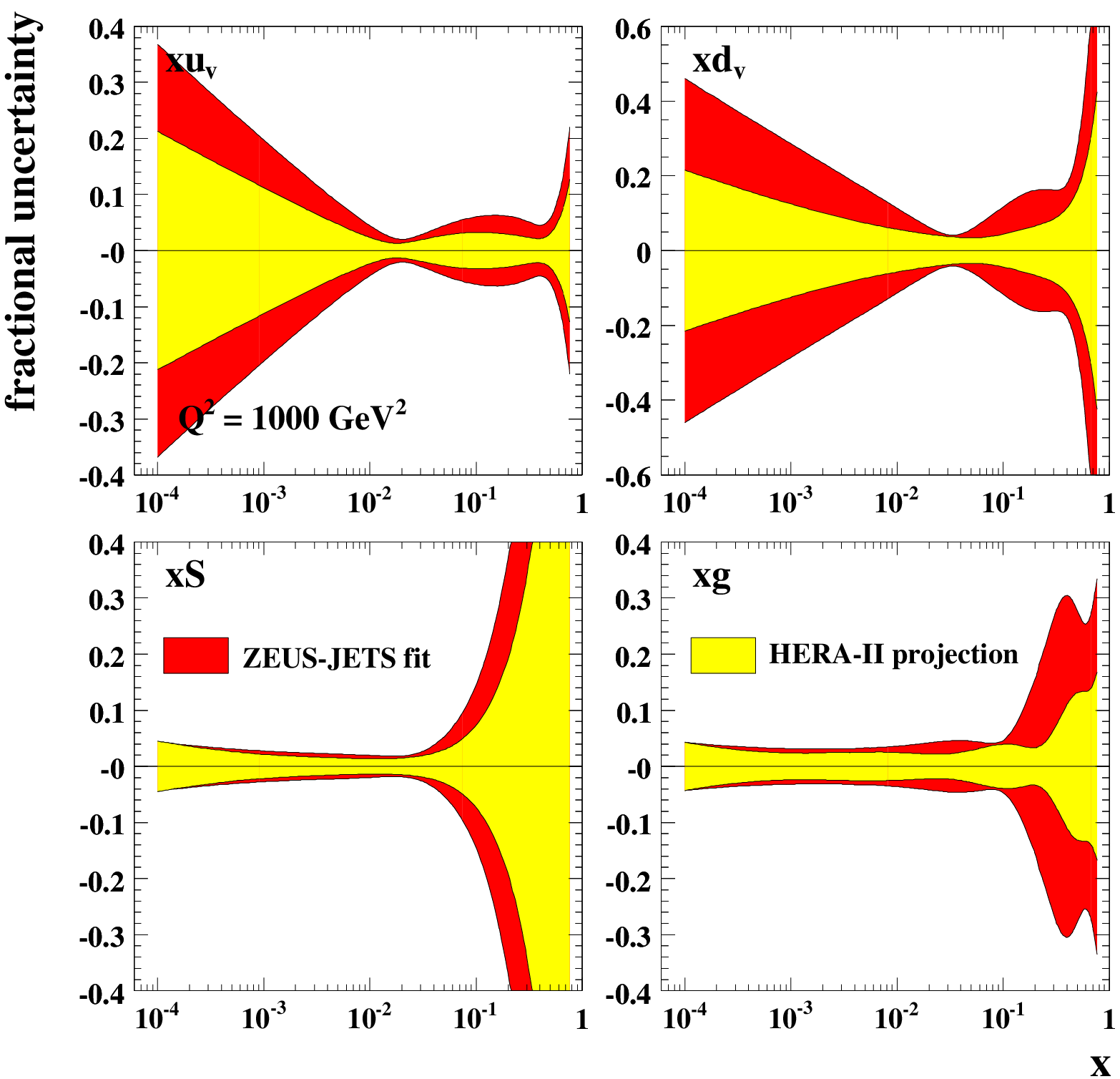} \\ 
      (a) & (b) \\ [0.4cm]
    \end{array}$
  \end{center}
  \vspace{-0.5cm}
  \caption{The ZEUS-pol PDFs for the valence ($u_{v}, d_{v}$), sea ($S$), and
    gluons ($g$) are shown in figure (a) compared with the ZEUS-JETS PDFs
    central values. 
    Figure (b) shows the uncertainties for the ZEUS-JETS PDFs in red and the
    HERA-II projected uncertainties in yellow. 
  }
  \label{fig-pdfs}
\end{figure}
\section{PDF Fits Using Only HERA Data and the Inclusion of Jet Data}
The PDFs are usually determined in global fits using data from many different
experiments. However, the high precision and wide kinematic coverage of 
existing HERA data allow precise extractions of the proton PDFs
using only HERA data. The use of HERA data alone eliminates the uncertainty
from heavy-target corrections and also avoids difficulties that can sometimes
arise from combining data sets from several different experiments.

The high statistics HERA NC data is used to determine the low $x$ sea and gluon
distributions  while information on the valence quarks is
provided by the higher-$Q^{2}$ NC and CC data. The gluon density contributes
indirectly to the inclusive DIS cross sections, however it makes a direct
contribution to the jet cross sections through boson-gluon fusion. 
The ZEUS collaboration has performed a combined NLO QCD fit (ZEUS-JETS
PDF~\cite{zjets}) to inclusive NC and CC DIS data as well as high precision
jet data in DIS and $\gamma p$ scattering.

The ZEUS-JETS PDFs agree well with the previous ZEUS-S PDF
global fits and are also compatible with the MRST~\cite{mrst} and
CTEQ~\cite{cteq} PDFs. The shapes of the PDFs are not changed significantly by
including jet data but the decrease in the uncertainty on the gluon
distribution is significant, approximately halved, in the mid-$x$ region over
the full $Q^{2}$ range. 

\section{Inclusion of New Data in PDFs and Future Projections from HERA}
The PDF uncertainties from current global fits are, in general, limited by
irreducible experimental systematics. In contrast, the fits to HERA data alone
are largely limited by the statistical precision of existing measurements. 
Since 2003, HERA has delivered a substantial amount of luminosity with
polarised lepton beams. Figure~\ref{fig-pdfs}(a) shows a new PDF fit named
ZEUS-pol~\cite{zpol} which includes HERA-II $e^{-}p$ NC and CC inclusive cross
section data with a total integrated luminosity of $121.5 \pbi$. This leads to
an improvement in PDF uncertainties at high $x$, especially for the
$u$ valence quark. 

As new HERA data is analysed, a significant impact on the gluon uncertainties
could be made by future jet cross section measurements in kinematic regions
optimised for PDF sensitivity. The effect on the PDF uncertainties using
the entire HERA data set has been estimated in the HERA-II
projection fit~\cite{gwen}.   
A total integrated luminosity of $700 \pbi$ was assumed for the high $Q^{2}$
inclusive data, and $500 \pbi$ was assumed for the jet measurements with
central values and systematic uncertainties taken from the published data
in each case. A set of optimised jet cross sections were included
for forward $\gamma p$ collisions assuming a luminosity of $500\pbi$.  

The increased statistical precision of the assumed amount of high $Q^{2}$ 
data  gives a significant improvement in the valence quark uncertainties over
the whole range of $x$. A significant improvement at high $x$ is seen for the
sea quarks, however the low $x$ sea and low $x$ gluons are not significantly
impacted as the data constraining this region tends to be at lower $Q^{2}$ and
so already systematically limited. Much improvement is seen in the
mid-to-high $x$ gluon which is constrained by jet data. 
Approximately half of the projected
reduction in the gluon uncertainties is due to the inclusion of optimised jet
cross sections.  

Accurate proton PDFs are of great importance, especially for the LHC proton-proton
collider which is planning to deliver high energy collisions in 2008.
With HERA shutting down in July 2007, the projected improvements to the 
PDF uncertainties using solely HERA data will be particularly relevant to
future physics at the LHC.

\end{document}